Title: Non-regular unary language and parallel communicating Watson-Crick automata systems

Authors: Kingshuk Chatterjee[1], Kumar Sankar Ray(corresponding author)[2]

Affiliations:
[1,2]Electronics and Communication Sciences Unit, Indian Statistical Institute, Kolkata-108

Address: Electronics and Communication Sciences Unit, Indian Statistical Institute, Kolkata-108.

Telephone Number: +918981074174

Fax Number:033-25776680

Email:ksray@isical.ac.in

# Non-regular unary language and parallel communicating Watson-Crick automata systems

*Kingshuk Chatterjee,[1] Kumar Sankar Ray[2]*

*Electronics and Communication Sciences unit, ISI, Kolkata.*

[1]kingshukchaterjee@gmail.com [2]ksray@isical.ac.in

*Abstract: In 2006, Czeizler et.al. introduced parallel communicating Watson-Crick automata system. They showed that parallel communicating Watson-Crick automata system can accept the non-regular unary language L={$a^{n^2}$, where $n > 1$} using non-injective complementarity relation and three components. In this paper, we improve on Czeizler et.al. work by showing that parallel communicating Watson-Crick automata system can accept the same language L using just two components.*

## I. INTRODUCTION

Martin-Vide et.al. [1] introduced parallel communicating automata systems. The system consists of many finite automata communicating with states. They also established that the computational power of such a system is equivalent to non-deterministic finite automata with multiple heads.

Watson-Crick automata are finite automata having two independent heads working on double strands where the characters on the corresponding positions of the two strands are connected by a complementarity relation similar to the Watson-Crick complementarity relation. The movement of the heads although independent of each other is controlled by a single state. Freund et.al.[2] introduced Watson-Crick automata. Its deterministic variants were introduced by Czeizler et.al. [3]. Work on state complexity of Watson-Crick automata are discussed in [4] and [5].

Czeizler et.al. introduced parallel Communicating Watson-Crick automata systems (PCWKS)[6] and further showed that with non-injective complementarity relation parallel communicating Watson-Crick automata system can accept the non-regular unary language L={$a^{n^2}$, where $n > 1$} with three components[7].

A parallel communicating Watson-Crick automata system [6] consists of several Watson-Crick automata each on its own input tape, and communicating by states. Every component of parallel communicating Watson-Crick automata system has its own double-stranded tape; the input is the same on all of them. An input is accepted by the system if all components are in final state and they completely parse the tape. Moreover, if one of the components stops before the others, the system halts and rejects the input.

In this paper we show that parallel communicating Watson-Crick automata system can accept the language L= {$a^{n^2}$, where $n > 1$} using just two components which is an improvement on the work by Czeizlers et. al. (see Section IV).

## II. BASIC TERMINOLOGY

The symbol V denotes a finite alphabet. The set of all finite words over V is denoted by $V^*$, which includes the empty word $\lambda$. The symbol $V^+=V^*- \{\lambda\}$ denotes the set of all non-empty words over the alphabet V. For $w \in V^*$, the length of w is denoted by $|w|$. Let $u \in V^*$ and $v \in V^*$ be two words and if there is some word $x \in V^*$, such that v=ux, then u is a prefix of v, denoted by $u \leq v$. Two words, u and v are prefix comparable denoted by $u \sim_p v$ if u is a prefix of v or vice versa.

A Watson-Crick automaton is a 6-tuple of the form $M=(V,\rho,Q,q_0,F,\delta)$ where V is an alphabet set, set of states is denoted by Q, $\rho \subseteq V \times V$ is the complementarity relation similar to Watson-Crick complementarity relation, $q_0$ is the initial state and $F \subseteq Q$ is the set of final states. The function $\delta$ contains a finite number of transition rules of the form $q \binom{w_1}{w_2} \to q'$, which denotes that the machine in state q parses $w_1$ in upper strand and $w_2$ in lower strand and goes to state q' where $w_1, w_2 \in V^*$. The symbol $\begin{bmatrix} w_1 \\ w_2 \end{bmatrix}$ is different from $\binom{w_1}{w_2}$. While $\binom{w_1}{w_2}$ is just a pair of strings written in that form instead of $(w_1, w_2)$, the symbol $\begin{bmatrix} w_1 \\ w_2 \end{bmatrix}$ denotes that the two strands are of same length i.e. $|w_1|=|w_2|$ and the corresponding symbols in two strands are complementarity in the sense given by the relation $\rho$. The symbol $\begin{bmatrix} V \\ V \end{bmatrix}_\rho = \{\begin{bmatrix} a \\ b \end{bmatrix} \mid a, b \in V, (a, b) \in \rho\}$ and $WK_\rho(V) = \begin{bmatrix} V \\ V \end{bmatrix}_\rho^*$ denotes the Watson-Crick domain associated with V and $\rho$.

A transition in a Watson-Crick finite automaton can be defined as follows:

For $\binom{x_1}{x_2}, \binom{u_1}{u_2}, \binom{w_1}{w_2} \in \binom{V^*}{V^*}$ such that $\begin{bmatrix} x_1 u_1 w_1 \\ x_2 u_2 w_2 \end{bmatrix} \in WK_\rho(V)$ and $q, q' \in Q$, $\binom{x_1}{x_2} q \binom{u_1}{u_2} \binom{w_1}{w_2} \Rightarrow \binom{x_1}{x_2} \binom{u_1}{u_2} q' \binom{w_1}{w_2}$ iff there is transition rule $q\binom{u_1}{u_2} \to q'$ in $\delta$ and $\overset{*}{\Rightarrow}$ denotes the transitive and reflexive closure of $\Rightarrow$. The language accepted by a Watson-Crick automaton M is $L(M) = \{w_1 \in V^* | q_0 \begin{bmatrix} w_1 \\ w_2 \end{bmatrix} \overset{*}{\Rightarrow} q \begin{bmatrix} \lambda \\ \lambda \end{bmatrix}$, with $q \in F$, $w_2 \in V^*$, $\begin{bmatrix} w_1 \\ w_2 \end{bmatrix} \in WK_\rho(V)\}$.

## III. PARALLEL COMMUNICATING WATSON-CRICK AUTOMATA SYSTEM

A parallel communicating Watson-Crick automata system of degree $n$, denoted by PCWK (n), is a $(n + 3)$-tuple

$A = (V, \rho, A_1, A_2, \ldots, A_n, K)$,

where

- $V$ is the input alphabet;
- $\rho$ is the complementarity relation;
- $A_i = (V, \rho, Q_i, q_i, F_i, \delta_i)$, $1 \leq i \leq n$, are Watson-Crick finite automata, where the sets $Q_i$ are not necessarily disjoint;
- $K = \{K_1, K_2, \ldots, K_n\} \subseteq \bigcup_{i=1}^{n} Q_i$ is the set of query states.

The automata $A_1, A_2, \ldots, A_n$ are called the *components* of the system A. Note that any Watson-Crick finite automaton is a parallel communicating Watson-Crick automata system of degree 1.

A configuration of a parallel communicating Watson-Crick automata system is a $2n$-tuple $(s_1, \binom{u_1}{v_1}, s_2, \binom{u_2}{v_2}, \ldots, s_n, \binom{u_n}{v_n})$ where $s_i$ is the current state of the component $i$ and $\binom{u_i}{v_i}$ is the part of the input word which has not been read yet by the component $i$, for all $1 \leq i \leq n$. We define a binary relation $\vdash$ on the set of all configurations by setting

$(s_1, \binom{u_1}{v_1}, s_2, \binom{u_2}{v_2}, \ldots, s_n, \binom{u_n}{v_n}) \vdash (r_1, \binom{u'_1}{v'_1}, r_2, \binom{u'_2}{v'_2}, \ldots, r_n, \binom{u'_n}{v'_n})$

if and only if one of the following two conditions holds:

- $K \cap \{s_1, s_2, \ldots, s_n\} = \emptyset$, $\binom{u_i}{v_i} = \binom{x_i}{y_i}\binom{u'_i}{v'_i}$, and $r_i \in \delta_i(s_i, \binom{x_i}{y_i})$, $1 \leq i \leq n$;

- for all $1 \leq i \leq n$ such that $s_i = K_{j_i}$ and $s_{j_i} \notin K$ we have $r_i = s_{j_i}$, whereas for all the other $1 \leq \ell \leq n$ we have $r_\ell = s_\ell$. In this case $\binom{u'_i}{v'_i} = \binom{u_i}{v_i}$, for all $1 \leq i \leq n$.

If we denote by $\vdash^*$ the reflexive and transitive closure of $\vdash$, then the language recognized by a PCWKS is defined as:

$L(\mathcal{A}) = \{w_1 \in V^* \mid (q_1, \begin{bmatrix} w_1 \\ w_2 \end{bmatrix}, q_2, \begin{bmatrix} w_1 \\ w_2 \end{bmatrix}, \ldots, q_n, \begin{bmatrix} w_1 \\ w_2 \end{bmatrix}) \vdash^* (s_1, \begin{bmatrix} \lambda \\ \lambda \end{bmatrix}, s_2, \begin{bmatrix} \lambda \\ \lambda \end{bmatrix}, \ldots, s_n, \begin{bmatrix} \lambda \\ \lambda \end{bmatrix}), s_i \in F_i, 1 \leq i \leq n\}$.

Intuitively, the language accepted by such a system consists of all words $w_1$ such that in every component we reach a final state after reading all input $\begin{bmatrix} w_1 \\ w_2 \end{bmatrix}$. Moreover, if one of the components stops before the others, the system halts and rejects the input. The above definition of parallel communicating Watson-Crick automata is in [6].

## IV. MAIN RESULT

In this Section, we design a parallel communicating Watson-Crick automata system with two components to accept the non-regular unary language $L = \{a^{n^2}, \text{where } n > 1\}$. In order to accept L, we use the same intuition as used by Czeizler et. al.[7]. If we consider the complementarity relation $\rho$ to be $\{(a,b), (a,c)\}$, then one of the many complementarity strings of a string in the language L must be of the form $b^n c^n b^n c^n \ldots$ where the total number of 'bc' and 'cb' pairs is (n-1). Thus, if we check for such a complementarity string in the lower strand then we will accept only those strings which are in the language. A string which is not in the language cannot have a complementarity string which is of the form $b^n c^n b^n c^n \ldots$ where the total number of 'bc' and 'cb' pairs is (n-1). For e.g. consider $a^{3^2}$ i.e. aaaaaaaaa where n=3 one of its many complementarity string is bbbcccbbb which is of the form $b^n c^n b^n c^n \ldots$ where the total number of 'bc' and 'cb' pairs is (n-1). Using the above stated idea we prove the following Theorem.

**Theorem 1:** A parallel communicating Watson-Crick automata system can accept the non-regular unary language, $L = \{a^{n^2}, \text{where } n > 1\}$, with just two components and non injective complementarity relation.

**Proof:** Let $A = (\{a,b,c\}, \rho, A_1, A_2, K)$ be a parallel communicating Watson-Crick automata system which accepts the language $L = \{a^{n^2}, \text{where } n > 1\}$ where $\rho = \{(a,b), (a,c)\}$, $K = \{K_1, K_2\}$ and $Q = \{q_0, q_1, q_2, q_3, q_4\}$.

The components of A are as follows:

$A_1 = (\ \{a,b,c\},\ K \cup Q \cup \{(q_0, \lambda, b), (q_0, \lambda, c), (q_1, aaa, \lambda), (q_2, a, c), (q_2, aa, b), (q_3, a, b), (q_3, aa, c), (q_2, \lambda, \lambda), (q_3, \lambda, \lambda), (q_4, \lambda, \lambda), s_2\},\ q_0,\ \{q_4\},\ \delta_1)$, and

$A_2 = (\ \{a,b,c\},\ K \cup Q \cup \{(q_0, \lambda, b), (q_0, \lambda, c), (q_1, aaa, \lambda), (q_2, a, c), (q_2, aa, b), (q_3, a, b), (q_3, aa, c), (q_2, \lambda, \lambda), (q_3, \lambda, \lambda), (q_4, \lambda, \lambda), (q_0, \lambda, b, \lambda), (q_0, \lambda, c, \lambda), (q_1, aaa, \lambda, b), (q_2, a, c, b), (q_2, aa, b, c), (q_3, a, b, c), (q_3, aa, c, b), (q_2, \lambda, \lambda, b), (q_3, \lambda, \lambda, c), (q_4, \lambda, \lambda, b), (q_4, \lambda, \lambda, c)\ \},\ q_0,\ \{q_4\},\ \delta_2)$, and

the transition functions of two components of A are defined in Table 1.

Table 1: Transition function of components of A

| Component $A_1$ | Component $A_2$ |
|---|---|
| $\delta_1(s_2, \binom{\lambda}{\lambda}) = K_2$ | $\delta_2(q, \binom{\lambda}{\lambda}) = K_1$ for all $q \in Q$ |
| $\delta_1(q_0, \binom{\lambda}{b}) = (q_0, \lambda, b)$ | $\delta_2((q_0, \lambda, b), \binom{\lambda}{\lambda}) = (q_0, \lambda, b, \lambda)$ |
| $\delta_1((q_0, \lambda, b), \binom{\lambda}{\lambda}) = s_2$ | $\delta_2((q_0, \lambda, b, \lambda), \binom{\lambda}{\lambda}) = q_0$ |
| $\delta_1(q_0, \binom{\lambda}{c}) = (q_0, \lambda, c)$ | $\delta_2((q_0, \lambda, c), \binom{\lambda}{\lambda}) = (q_0, \lambda, c, \lambda)$ |
| $\delta_1((q_0, \lambda, c), \binom{\lambda}{\lambda}) = s_2$ | $\delta_2((q_0, \lambda, c, \lambda), \binom{\lambda}{\lambda}) = q_1$ |
| $\delta_1(q_1, \binom{aaa}{\lambda}) = (q_1, aaa, \lambda)$ | $\delta_2((q_1, aaa, \lambda), \binom{a}{b}) = (q_1, aaa, \lambda, b)$ |
| $\delta_1((q_1, aaa, \lambda), \binom{\lambda}{\lambda}) = s_2$ | $\delta_2((q_1, aaa, \lambda, b), \binom{\lambda}{\lambda}) = q_2$ |
| $\delta_1(q_2, \binom{a}{c}) = (q_2, a, c)$ | $\delta_2((q_2, a, c), \binom{a}{b}) = (q_2, a, c, b)$ |
| $\delta_1((q_2, a, c), \binom{\lambda}{\lambda}) = s_2$ | $\delta_2((q_2, a, c, b), \binom{\lambda}{\lambda}) = q_2$ |
| $\delta_1(q_2, \binom{aa}{b}) = (q_2, aa, b)$ | $\delta_2((q_2, aa, b), \binom{a}{c}) = (q_2, aa, b, c)$ |
| $\delta_1((q_2, aa, b), \binom{\lambda}{\lambda}) = s_2$ | $\delta_2((q_2, aa, b, c), \binom{\lambda}{\lambda}) = q_3$ |
| $\delta_1(q_3, \binom{a}{b}) = (q_3, a, b)$ | $\delta_2((q_3, a, b), \binom{a}{c}) = (q_3, a, b, c)$ |
| $\delta_1((q_3, a, b), \binom{\lambda}{\lambda}) = s_2$ | $\delta_2((q_3, a, b, c), \binom{\lambda}{\lambda}) = s_3$ |
| $\delta_1(q_3, \binom{aa}{c}) = (q_3, aa, c)$ | $\delta_2((q_3, aa, c), \binom{a}{b}) = (q_3, aa, c, b)$ |
| $\delta_1((q_3, aa, c), \binom{\lambda}{\lambda}) = s_2$ | $\delta_2((q_3, aa, c, b), \binom{\lambda}{\lambda}) = q_2$ |
| $\delta_1(q_2, \binom{\lambda}{\lambda}) = (q_2, \lambda, \lambda)$ | $\delta_2((q_2, \lambda, \lambda), \binom{a}{b}) = (q_2, \lambda, \lambda, b)$ |
| $\delta_1((q_2, \lambda, \lambda), \binom{\lambda}{\lambda}) = s_2$ | $\delta_2((q_2, \lambda, \lambda, b), \binom{\lambda}{\lambda}) = q_4$ |
| $\delta_1(q_3, \binom{\lambda}{\lambda}) = (q_3, \lambda, \lambda)$ | $\delta_2((q_3, \lambda, \lambda), \binom{a}{c}) = (q_3, \lambda, \lambda, c)$ |
| $\delta_1((q_3, \lambda, \lambda), \binom{\lambda}{\lambda}) = s_2$ | $\delta_2((q_3, \lambda, \lambda, c), \binom{\lambda}{\lambda}) = q_4$ |
| $\delta_1(q_4, \binom{\lambda}{\lambda}) = (q_4, \lambda, \lambda)$ | $\delta_2((q_4, \lambda, \lambda), \binom{a}{b}) = (q_4, \lambda, \lambda, b)$ |
| $\delta_1((q_4, \lambda, \lambda), \binom{\lambda}{\lambda}) = s_2$ | $\delta_2((q_4, \lambda, \lambda, b), \binom{\lambda}{\lambda}) = q_4$ |
| | $\delta_2((q_4, \lambda, \lambda), \binom{a}{c}) = (q_3, \lambda, \lambda, c)$ |
| | $\delta_2((q_4, \lambda, \lambda, c), \binom{\lambda}{\lambda}) = q_4$ |

The parallel communicating Watson-Crick automata system A works in the following manner:

We use the lower head of the first component $A_1$ of A and the lower head of the second component $A_2$ of A to check whether the lower strand have alternative blocks of b's and c's which are of equal size.

We employ the upper and lower head of the first component $A_1$ of A to check whether the number of 'bc' and 'cb' pairs in the lower strand is one less than the total number of b's in the first block.

Parallel communicating Watson-Crick automata system A moves the lower head of the first component $A_1$ until it comes across the first 'c' in the lower strand. If the lower strand is of the form $b^n c^n b^n c^n$..., then when the lower head reads the first 'c' the upper head is n symbols behind the lower head (as the first 'c' comes after n b's). When $A_1$ comes across the first 'c' in its lower head, its upper head reads three a's then the control switches to the second component $A_2$ and $A_2$'s upper head reads 'a' and its lower head reads 'b'. Every time the lower head of the second component $A_2$ reads a character the upper head of the second component $A_2$ reads an 'a'(we do not require the upper head of second component for any computation). This is done to ensure both heads of second component reach the end of their tape at the same time. Then the control again switches back to component $A_1$. After that, every time the lower head of $A_1$ reads a 'b'/'c' the upper head of $A_1$ reads an 'a' and the control switches to component $A_2$ where the lower head of component $A_2$ reads a 'c/b' following which the control again switches back to component $A_1$. As the lower head of $A_2$ was at the beginning of the tape when lower head of $A_1$ read the first 'c' if the input is of the form $b^n c^n b^n c^n$... then for every 'b'/ 'c' read by the lower head of $A_1$ the lower head of $A_2$ will read a 'c'/'b'. This ensures that the lower strand have alternate blocks of 'b' s and 'c' s of equal size. If the lower strand does not have equal blocks of 'b' and 'c' then the second component will halt without both its heads reaching the end of the tape. Thus, $A_2$ rejects the input.

Every time there is a change of symbol read by the lower head of $A_1$ from 'b' to 'c' or from 'c' to 'b' the upper head of $A_1$ reads an extra 'a' in addition to the steps mentioned above. Thus, for each 'bc'or 'cb' pair except for the first 'bc' pair the upper head of $A_1$ reads an extra 'a'. For the first 'bc' pair, the upper head of $A_1$ reads two extra a's. The extra a's read by the upper head of $A_1$ for each 'bc' or 'cb' pair $A_1$ comes across in its lower strand enables the upper head of $A_1$ which was n symbols behind the lower head to catch up with lower head and reach the end of the tape.

Therefore the two heads of component $A_1$ reach the end of their respective tapes only when the number of 'bc' or 'cb' pairs is (n-1). If the total number of 'bc' and 'cb' pairs is more than (n-1) the upper head of $A_1$ will reach the end of the tape earlier than the lower head. As a result, when lower head of $A_1$ reads a character in its lower strand there will be no 'a' available in the upper head to read thus the automaton $A_1$ will halt with its upper head at the end of the tape and lower hand not at the end. If the total number of 'bc' and 'cb' pairs is less than (n-1) the lower head will finish earlier than upper head and thus the automaton $A_1$ will halt with its upper head still needing to consume 'a'. As there are no transitions in $A_1$ where the upper head can read 'a' without the lower head reading any character, so upper head of $A_1$ will not reach the end of the tape. Both the heads of $A_1$ reach the end of their respective tapes only if the number of 'bc' or 'cb' pairs is one less than the size of the first block of 'b's.

From the above explained structure of A, we see that the upper head and lower head of both the components of A will reach the end of their respective tapes only when the upper strand of input to A contains only 'a's and lower strand is of the form $b^n c^n b^n c^n$... where the number of 'bc' and 'cb' pairs are (n-1).

Thus, if we analyze the string the parallel communicating Watson-Crick automata system A accepts we see that it accepts all those strings composed of letter 'a' which has a complementarity string in the form $b^n c^n b^n c^n$... where the total number of 'bc' and 'cb' pairs is (n-1).

Consider a string w in L=$\{a^{n^2},$ where n > 1$\}$, one of its many complementarity strings must be of the form $b^n c^n b^n c^n$... where the number of 'bc' and 'cb' pairs is (n-1), e.g. $a^{3^2}$; i.e. aaaaaaaaa where n=3 one of its many complementarity strings is bbbcccbbb. Therefore w is accepted by A.

Now consider a string w not in L, no matter what complementarity string of w we take it can never be of the form $b^n c^n b^n c^n$... where the number of 'bc' and 'cb' pairs is (n-1). So w will not be accepted by A.

Thus, we can say that A accepts L.

**Lemma 1:** Non-deterministic multi-head finite automata cannot accept non-regular unary languages.

The proof of this Theorem is in [7].

**Corollary 1:** Parallel communicating Watson-Crick automata system with just two components accepts a language which is not accepted by any multi-head finite automata.

**Proof:** The proof of this Corollary follows from Theorem 1 and Lemma 1. From Theorem 1 we see that there exists a parallel communicating Watson-Crick automata system that can accept the non-regular unary language, L=$\{a^{n^2},$ where n > 1$\}$, with just two components and non injective complementarity relation and from Lemma 1 we know that non-deterministic multi-head finite automata cannot accept non-regular unary languages. Thus we conclude from Theorem 1 and Lemma 1 that parallel communicating Watson-Crick automata system with just two components accepts a language which is not accepted by any multi-head finite automata.

## V. CONCLUSION

In this paper, we show that parallel communicating Watson-Crick automata system can accept the non-regular unary language L={$a^{n^2}$, where n > 1} using non-injective complementarity relation and just two components. We use one component less than Czeizler et. al. to accept the same language. Our result also enables us to show that with just two components parallel communicating Watson-Crick automata system accepts a language which is not accepted by any multi-head finite automata.